\documentclass[aps,prl,twocolumn,groupedaddress,showpacs,floatfix]{revtex4}
\usepackage[]{graphicx}
\usepackage[]{subfigure}
\begin{document}
\title{Bright source of spectrally uncorrelated polarization-entangled photons with nearly single-mode emission}
\author{P. G. Evans}
\email[]{evanspg@ornl.gov}
\affiliation{Center for Quantum Information Science, Computing and Computational Sciences Directorate, Oak Ridge National Laboratory, Oak Ridge, Tennessee 37831, USA}
\author{J. Schaake}
\affiliation{Department of Physics and Astronomy, University of Tennessee, Knoxville, Tennessee 37996, USA}
\author{R. S. Bennink}
\affiliation{Center for Quantum Information Science, Computing and Computational Sciences Directorate, Oak Ridge National Laboratory, Oak Ridge, Tennessee 37831, USA}
\author{W. P. Grice}
\affiliation{Center for Quantum Information Science, Computing and Computational Sciences Directorate, Oak Ridge National Laboratory, Oak Ridge, Tennessee 37831, USA}
\author{T. S. Humble}
\affiliation{Center for Quantum Information Science, Computing and Computational Sciences Directorate, Oak Ridge National Laboratory, Oak Ridge, Tennessee 37831, USA}
\date{December 13, 2010}
\begin{abstract}
We present results of a bright polarization-entangled photon source operating at 1552 nm via type-II collinear degenerate spontaneous parametric down-conversion in a periodically poled potassium titanyl phosphate crystal. We report a conservative inferred pair generation rate of 123,000 pairs/s/mW into collection modes. Minimization of spectral and spatial entanglement was achieved by group velocity matching the pump, signal and idler modes and through properly focusing the pump beam. By utilizing a pair of calcite beam displacers, we are able to overlap photons from adjacent down-conversion processes to obtain polarization-entanglement visibility of 94.7 +/- 1.1\% with accidentals subtracted.\end{abstract}
\maketitle
Spontaneous parametric down-conversion (SPDC) is the leading mechanism for realizing photonic quantum states. For many applications, particularly those that involve the interference of photons from multiple SPDC sources, the individual photon pairs are required to be spectrally and spatially pure. This is because multimode down-converted photons carry distinguishing information that undermines the interference central to such applications \cite{pittman2003}. It has been shown previously that spectral entanglement can be eliminated through source engineering. This approach has led to an experimental demonstration of heralded generation of spectrally pure single photons from type-II SPDC \cite{mosley2008, migdall2010}, and has also inspired similar approaches of type-I SPDC \cite{vicent2010, uren2005} and spontaneous four-wave mixing in fibers \cite{soller2010, cohen2009, halder2009}.

In this Letter, we demonstrate the first bright SPDC source for polarization-entangled biphoton states that are uncorrelated in both the spectral and spatial degrees of freedom. As in the previous works, our source is engineered for spectral purity through the selection of the pump properties and the phase-matching characteristics of the SPDC medium. In addition, special attention is paid to the spatial entanglement. Specifically, our source is based on collinear type-II SPDC in noncritically phase-matched periodically poled potassium titanyl phosphate (PPKTP), a geometry that eliminates walk-off effects and maximizes the overlap of the pump, signal, and idler fields \cite{fedrizzi2007}. In this configuration, a properly focused pump causes nearly all of the photon to be emitted into a single spatial mode \cite{grice2010, bennink2010}. The combination of these techniques for controlling the SPDC output gives rise to a single-mode emission rate of 123,000 pairs/s/mW. The source is further configured in a novel arrangement of calcite beam displacers that separate and then recombine photons generated by either of the two parallel pump beams.

The spectral and spatial properties of SPDC photons are determined by the pump field and by the dispersive properties of the crystal. The two-photon probability amplitude is the product of a pump function and a phase-matching function. Entanglement can be eliminated only if this product yields no correlations between the signal and idler photon properties. By careful selection of the wavelength, pulse duration, and focus of the pump, as well as the crystal material and length, it is possible to minimize entanglement in these degrees of freedom. In general, the spectral and spatial properties are not independent, but for the purposes of illustrating the design principles, it is sufficient here to treat them separately.

To eliminate spectral entanglement, the shapes of the pump function and phase-matching function must be chosen correctly \cite{grice2001}. The pump function, which describes the range of energies available for down-conversion, generally yields negatively correlated photon energies --- the signal and idler energies must sum to an energy somewhere in the pump spectrum, so a longer wavelength for one photon is necessarily accompanied by a shorter wavelength for the other. This correlation is strongest for a monochromatic pump, for which there is but a single pump energy. Therefore, one requirement for the elimination of spectral entanglement is a broad pump spectrum. However, this alone will not eliminate spectral entanglement if the phase-matching function also leads to negatively correlated energies. 

Whereas the pump function describes the range of energies available for down-conversion, the phase-matching function describes the ways that the pump energies may be distributed to the signal and idler photons. The influence of the phase-matching function on the spectral properties of the photons is revealed by noting that the function has appreciable value only for $\Delta{k}L \simeq 0$, where $L$ is the crystal length and where $\Delta{k}=k_{p}-k_{s}-k_{i}$ is the wavevector mismatch. Using the approximation $k\simeq k_0+\nu k^{\prime}$ and the fact that $k_{p0}-k_{s0}-k_{i0}=0$ for a phase-matched interaction, we have $\Delta{k}\simeq(\nu_{s}+\nu_{i})k_{p}^{\prime}-\nu_{s}k_{s}^{\prime}-\nu_{i}k_{i}^{\prime}$. Here $\nu=\omega-\omega_0$ and $k^{\prime} = \partial{k}/\partial{\omega}$. Imposing the requirement that $\Delta{k}\simeq0$ yields
\begin{equation}
\nu_{s}=-\nu_{i}\frac{k_{p}^{\prime}-k_{i}^{\prime}}{k_{p}^{\prime}-k_{s}^{\prime}}
\label{eqn:groupvelmatch}
\end{equation}
such that the phase-matching function leads to positively correlated photon energies only if $k_{p}^{\prime}$ lies between $k_{s}^{\prime}$ and $k_{i}^{\prime}$ or, since the group velocity $v_{g} = 1/k^{\prime}$, if the group velocity of the pump lies between the group velocities of the signal and idler photons.

It is difficult to satisfy the group velocity matching condition as defined in Eq. (\ref{eqn:groupvelmatch}) for visible wavelengths with most materials since normal material dispersion results in lower group velocities for the bluer pump wavelengths. However, dispersion is more accommodating at longer wavelengths, and solutions can be found for several type-II materials \cite{grice2001}. In particular, group velocity matching can be achieved with type-II SPDC in KTP with a pump wavelength range of 650-900 nm for degenerate down-conversion to 1.3-1.8 $\mu$m. Once the material and pump wavelength have been specified, the widths of the pump and phase-matching functions must be chosen so that the resulting probability amplitude exhibits neither positive nor negative correlations in the photon energies. This requirement leads to a specific relationship between the pump bandwidth and the crystal length. For the 20-mm PPKTP crystal used in our source, our calculations predict that the spectral entanglement will be minimized, with a spectral Schmidt number of 1.06, using a 776 nm pump with a bandwidth corresponding to a transform-limited pulse duration of 1.3 ps. We note that, with a longer crystal and/or a broader pump bandwidth, it is possible to generate photon pairs having positive energy correlations, as has been shown previously \cite{shimizu2009}.

The factors that must be considered to minimize the spatial (transverse momentum) entanglement are similar to those pertaining to the spectral entanglement. As in the spectral domain, the pump function leads to a tendency toward negatively correlated transverse momenta --- the signal and idler momenta must sum to a momentum somewhere in the pump spectrum, so the emission directions of the photons are negatively correlated. This correlation is strongest when the pump transverse momentum spectrum is narrow, i.e., when the pump is collimated. A necessary condition for the elimination of spatial entanglement, therefore, is a pump with a broad transverse momentum spectrum, a requirement that is easily met by focusing. 

The requirement that the phase-matching function yield toward positively correlated transverse momenta is satisfied in most down-conversion materials, particularly when there is not spatial walk-off. This is the case for the noncritically phase-matched PPKTP. All that remains then is to choose the widths of the pump and phase-matching functions so as to eliminate the spatial entanglement. For the 20-mm PPKTP crystal used in our source, calculations predict the best performance with the pump having a divergence of 13.1 mrad and focused at the center of the crystal. It is shown elsewhere \cite{grice2010} that the conditions that minimize spatial entanglement are the same conditions that maximize coupling to single-mode collection optics. In our case, the signal and idler photons are predicted to be emitted collinearly into single modes with divergences of 18.4 and 18.1 mrad, respectively.

Our approach provides several advantages in comparison to traditional multiphoton entanglement experiments: namely
\begin{itemize}
\item the pump wavelength of 776 nm is accessible with a tunable pulsed Ti:Sapphire laser, without the requirement for a second-harmonic generation crystal to double the pump wavelength as in the usual UV $\rightarrow$ Vis down-conversion schemes,
\item the minimization of spectral and spatial entanglement by the source removes any need for interference and spatial filters to be used, with SPDC photons emitted into a single spatial mode for optimal coupling to collection optics and
\item 776 nm $\rightarrow$ 1552 nm SPDC occurs in the technologically important telecom band where optical fibers exhibit minimal attenuation and standard telecoms equipment is readily available.
\end{itemize}

\begin{figure}
  \subfigure[Experimental setup]{\includegraphics[scale=0.3]{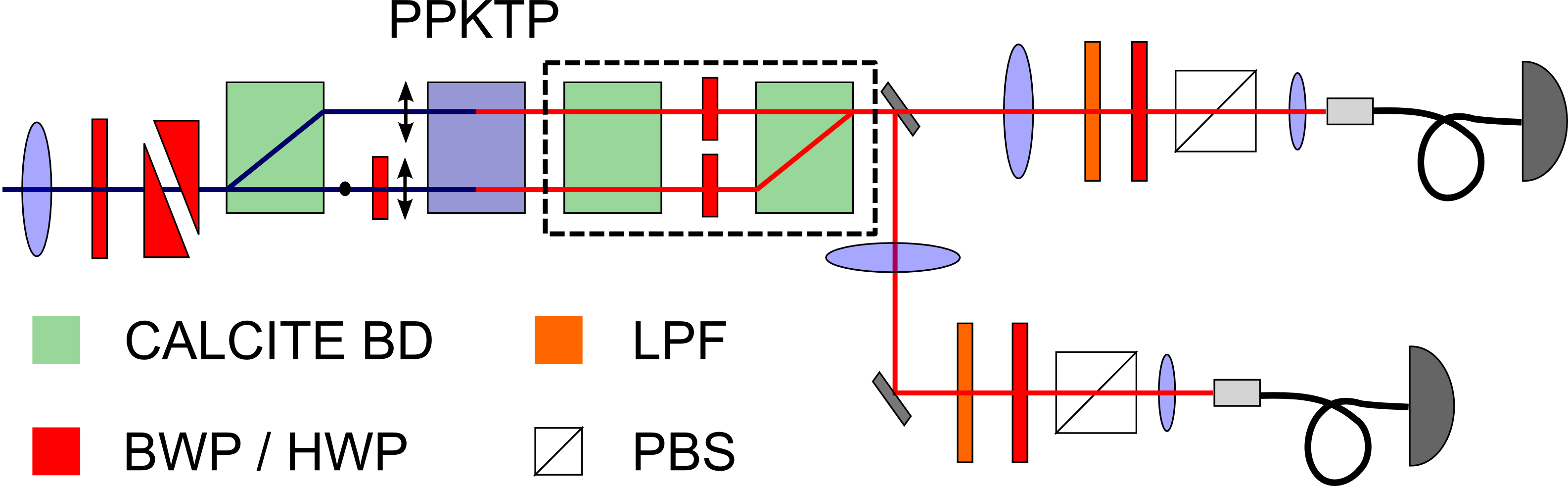}}
  \subfigure[Detail of beam displacer pair inset]{\includegraphics[scale=0.3]{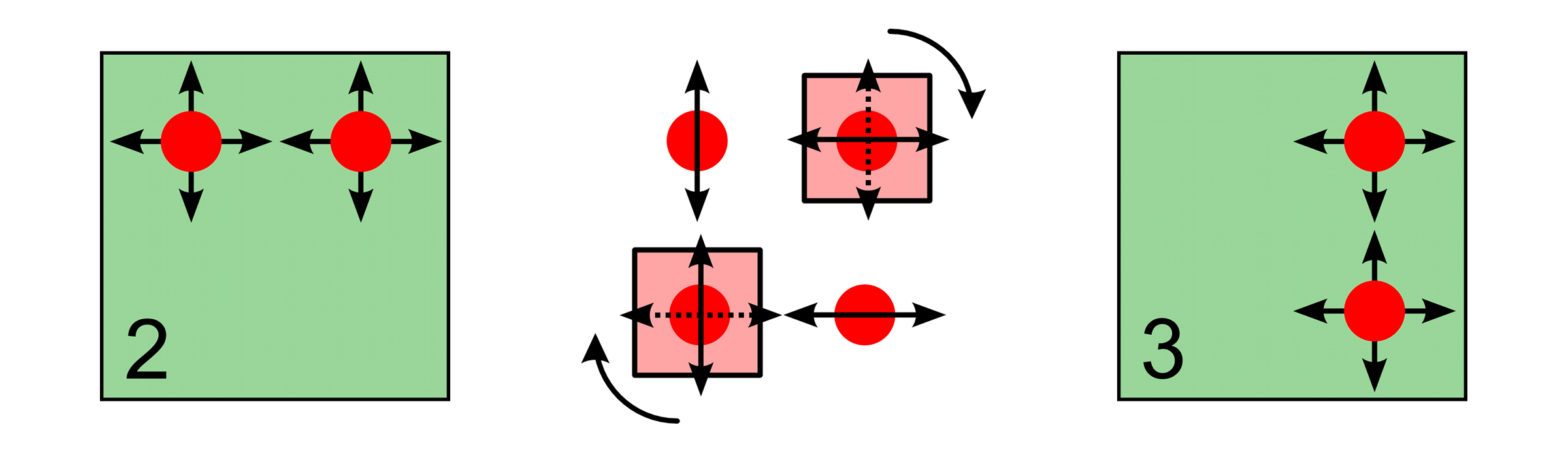}}
  \caption{(Color online) (a) Experimental setup. (b) Detail of the BD pair showing the beam configuration on entry to BD2 (left diagram), the action of the half-wave plates between BD2 and BD3 (middle diagram), and upon exit from BD3 (right diagram). LPF: long-pass filter; PBS: polarization beam splitter.}
  \label{fig:layout}
\end{figure}

The experimental setup, which is a modification of the scheme first presented in \cite{fiorentino2008}, is illustrated in Fig. \ref{fig:layout}. A 776 nm pump beam from a Coherent Mira Ti:Sapphire laser is incident upon a lens, half-wave plate (HWP1), birefringent wedge pair (BWP), and beam displacer (BD1). BD1 displaces the orthogonal polarizations of the pump components by 4.2 mm; the vertically polarized pump component is passed through the half-wave plate HWP2 oriented with the fast axis at 45$^{\circ}$. The two pump beams incident on the PPKTP crystal are horizontally polarized with a focus of 13.1 mrad to satisfy type-II phase matching and to minimize spatial entanglement as described above. Both pump waists are located midway along the length of the PPKTP crystal.

Upon emergence from the PPKTP crystal, the signal and idler photons from the two down-conversion regions are incident upon beam displacer BD2, which acts to displace the signal and idler photons vertically, creating four beam paths, shown in detail in Fig. \ref{fig:layout}b. Thin half-wave plates rotate the photon polarizations in two of the paths before all four beams are incident on beam displacer BD3, which acts to recombine photons such that both signal photons are emergent in the upper path and both idler photons are emergent in the lower path. Lenses following the beam displacers are used to match the spatial mode of the signal and idler photons to the collection optics. A pair of long-pass filters remove the residual pump from each arm and the photons are coupled into single-mode fibers. Wave plates and polarizers are placed in the two beams for analysis of the polarization-entangled state.

Photon detection is accomplished using two fiber-coupled idQuantique id200 InGaAs/InP avalanche photodiodes with a reported detection efficiency of approximately 10\%. Both detectors are set to use a 2.5 ns gate width and are triggered at 4.75 MHz synchronized with the 76 MHz pulse train from the Ti:Sapphire. Output pulses from the id200s, corresponding to photon detection events, are input into ORTEC counting, delay and coincidence logic circuits for computer readout.

\begin{figure}
  \includegraphics[scale=0.4]{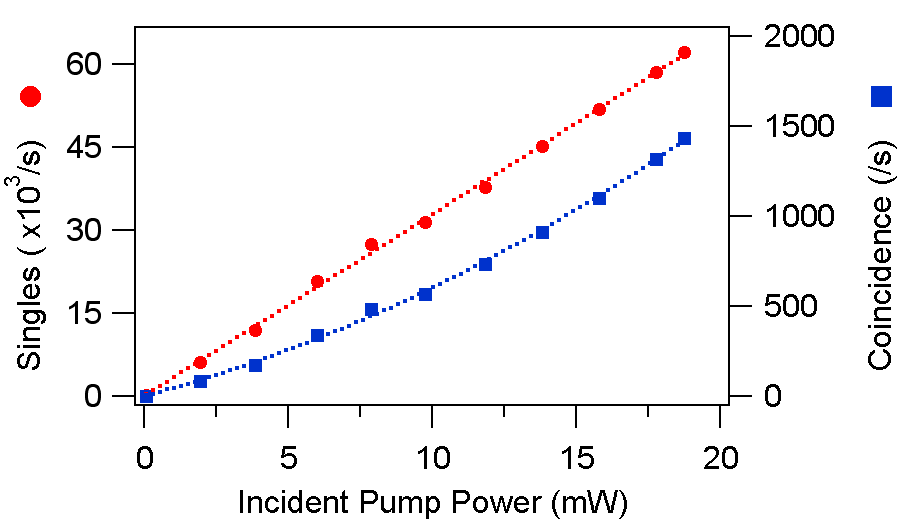} 
  \caption{(Color online) Singles $\sqrt{A.B}$ (left axis) and coincidence counts (right axis) vs incident pump power. The dashed lines serve as guides to the eye.}
  \label{fig:counts_vs_pump}
\end{figure}

The singles and coincidence counts are measured as a function of incident pump power and are displayed in Fig. \ref{fig:counts_vs_pump}. The half-wave plates and polarization beam splitters were removed from both collection arms for this measurement. The single counts are linear with respect to increasing pump power; the coincidence counts exhibit a slight nonlinear trend that we attribute to multiple-pair generation at higher pump power \cite{wong2008}. By taking into account Fresnel losses from the uncoated PPKTP crystal and calcite beam displacers, plus attenuation by the pair of long-pass filters in each arm, we estimate 38\% transmission at 1552 nm through the system. In conjunction with a conservative 50\% free-space to SMF coupling efficiency and 10\% detection efficiency for each id200 detector, our detection efficiency for 1552 nm photons is 1.9\%. Given the actual pair generation rate is equal to the measured coincidence rate divided by the singles detection efficiency squared, we conservatively infer the pair generation rate of our source to be 123,000 pairs/s/mW pump. To the best of our knowledge, this is the brightest narrow-band source operating in the telecom band reported to date.

\begin{figure}
  \includegraphics[scale=0.4]{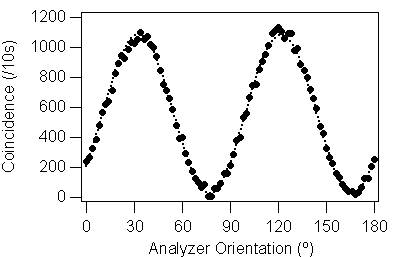} 
  \caption{Coincidence counts vs analyzer orientation angle in the $\pm$45$^{\circ}$ polarization basis showing a visibility of 94.7 $\pm$ 1.1\%. The incident pump power is 16 mW and accidental counts have been subtracted.}
  \label{fig:good_visibility}
\end{figure}

Polarization correlation measurements were carried out with the half-wave plates and polarization beam splitters reinserted in both arms. Figure \ref{fig:good_visibility} shows polarization correlation plots in the $\pm$45$^{\circ}$ basis. The incident pump power was set to 16 mW and accidentals were subtracted. Curve fitting to the data yields a visibility of 94.7 $\pm$ 1.1\%.

\begin{figure}
  \includegraphics[scale=0.4]{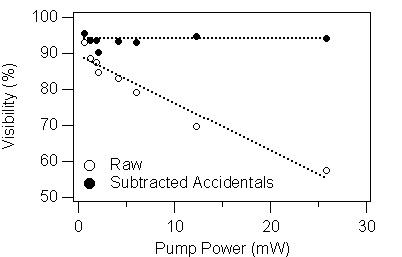} 
  \caption{Raw ($\circ$) and subtracted accidentals ($\bullet$) visibility in the $\pm$45$^{\circ}$ polarization basis as a function of incident pump power. The dashed lines indicate linear fits to the data.}
  \label{fig:vis_vs_pump}
\end{figure}

In order to examine the effect of multipair generation on the $\pm$45$^{\circ}$ basis visibility, we conducted several polarization correlation measurements with various incident pump powers. The results are presented in Fig. \ref{fig:vis_vs_pump}. The open circles show the uncorrected data and exhibit a linear decrease with respect to increasing pump power, providing evidence of multipair generation at higher pump power. The filled circles represent the corrected data, i.e., raw data with accidentals subtracted, which average 94\% and are constant with respect to incident pump power as one would expect. We considered several explanations that would lead to our visibility being capped to 94\%: differences in the optical path lengths of the two down-conversion processes, small differences in signal and idler photon wavelengths due to poling inhomogeneities, and errors in wave plate orientation. Optical path differences were minimized by positioning the birefringent wedge pair BWP to maximize visibility. We measured the spectra of all four photons using a monochromator and found that any wavelength differences were on the order of the monochromator resolution. Thus we conclude that the likely source of the reduced visibility is orientation error for the two wave plates placed between BD2 and BD3.

\begin{figure}
  \subfigure[1.3 ps pump]{\includegraphics[scale=0.4]{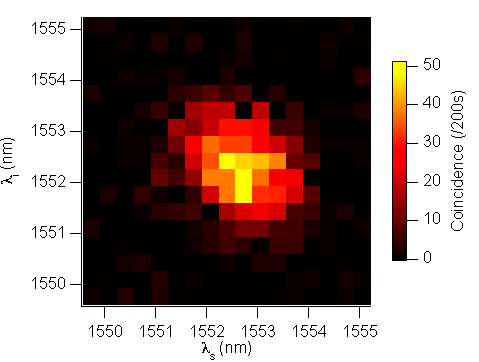}\label{fig:JSAgood}}
  \subfigure[1.9 ps pump]{\includegraphics[scale=0.4]{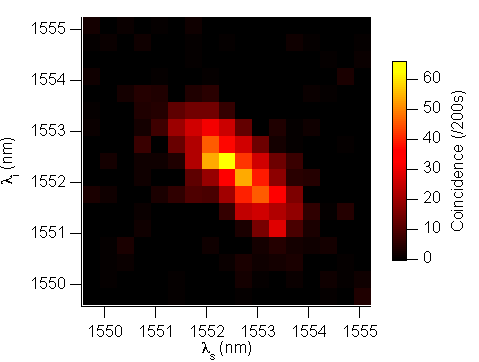}\label{fig:JSAok}}
\caption{(Color online) Joint spectral intensities using two pump pulse durations, 1.3 ps (a) and 1.9 ps (b). Analysis yields spectral Schmidt numbers of 1.07 and 1.16, respectively.}
\label{fig:JSA}
\end{figure}

We performed joint spectral intensity measurements on our source using a customized dual-slit scanning monochromator with 0.3 nm resolution. By controlling the pump laser mode locking, we were able to adjust the pump pulse duration and measure the joint spectral intensity at several different pump pulse durations. Figure \ref{fig:JSA} shows the measured joint spectral intensity taken with an incident pump power of 25 mW and 200 s counts per measurement and with accidental coincidences subtracted, for an optimal pump laser pulse duration of 1.3 ps [Fig. \ref{fig:JSAgood}] and 1.9 ps [Fig. \ref{fig:JSAok}] respectively. Analysis of the raw data yields spectral Schmidt numbers of 1.07 and 1.16, respectively, in agreement to 1.06 as predicted by our calculations for the optimal case. Furthermore, changes in the joint spectral intensity can clearly be seen between the different pump pulse durations.

In summary, we have demonstrated a unique source insofar as entanglement in the spatial and spectral degrees of freedom has been minimized by appropriate pump focusing and careful selection of pump bandwidth, wavelength, and phase matching with type-II PPKTP. As a result, photons are emitted into single spectral and spatial modes which require no spectral or spatial filtering to observe high multiphoton visibility of 94.7 $\pm$ 1.1\% with an inferred pair generation rate of 123,000/s/mW pump. By using a novel arrangement of beam displacers to recombine the signal and idler photons into distinct paths, we have demonstrated a source design that can be extended to the construction of multiphoton ($\geq$ 4) polarization-entangled states.

We thank M. Fiorentino and A. Migdall for enlightening discussions. Research is sponsored by the Laboratory Directed Research and Development Program of Oak Ridge National Laboratory, managed by UT-Battelle, LLC for the U. S. Department of Energy under Contract No. De-AC05-00OR22725.


\begin{thebibliography}{99}  
\bibitem{pittman2003}T. B. Pittman et al., IEEE J. Sel. Topics in Quantum Electron. \textbf{9}, 1478 (2003).
\bibitem{mosley2008}P. J. Mosley et al., Phys. Rev. Lett. \textbf{100}, 133601 (2008).
\bibitem{migdall2010}Z. H. Levine et al., Opt. Express \textbf{18}, 4, 3708-3718 (2010).
\bibitem{uren2005}A. B. U'Ren et al., Laser Phys. \textbf{15}, 146 (2005).
\bibitem{vicent2010}L. E. Vicent et al., New J. Phys. \textbf{12}, 093027 (2010).
\bibitem{soller2010} C. S\"oller et al., Phys. Rev. A \textbf{81}, 031801(R) (2010).
\bibitem{cohen2009} O. Cohen et al., Phys. Rev. Lett. \textbf{102}, 123603 (2009).
\bibitem{halder2009} M. Halder et al., Opt. Express \textbf{17}, 4670-4676 (2009).
\bibitem{fedrizzi2007} A. Fedrizzi et al., Opt. Express \textbf{15}, 23, 15377-15386 (2007).
\bibitem{grice2010}W. P. Grice et al., Submitted to Phys. Rev. A.
\bibitem{bennink2010}R. S. Bennink, Phys. Rev. A \textbf{81}, 053805 (2010).
\bibitem{grice2001}W. P. Grice, A. B. U'Ren, and I. A. Walmsley, Phys. Rev. A \textbf{64}, 063815 (2001).
\bibitem{shimizu2009} R. Shimizu, and K. Edamatsu, Opt. Express \textbf{17}, 16385-16393 (2009).
\bibitem{fiorentino2008}M. Fiorentino and R. Beausoleil, Opt. Express \textbf{16}, 20149 (2008).
\bibitem{wong2008} O. Kuzucu and F. N. C. Wong, Phys. Rev. A \textbf{77}, 032314 (2008).
\end{thebibliography}
\end{document}